# Mesoscale microscopy for micromammals: image analysis tools for understanding the rodent brain

Authors
Adam L . Tyson, Troy W. Margrie*

Affiliations
Sainsbury Wellcome Centre, University College London, 25 Howland Street, London, W1T 4JG , United Kingdom

*t.margrie@ucl.ac.uk

**Abstract**

Over the last ten years, developments in whole-brain microscopy now allow for high-resolution imaging of intact brains of small rodents such as mice. These complex images contain a wealth of information, but many neuroscience laboratories do not have all of the computational knowledge and tools needed to process these data. We review recent open source tools for registration of images to atlases, and the segmentation, visualisation and analysis of brain regions and labelled structures such as neurons. Since the field lacks fully integrated analysis pipelines for all types of whole-brain microscopy analysis, we propose a pathway for tool developers to work together to meet this challenge.

## 1. Mesoscale whole brain imaging

Developing a deeper understanding of the brain requires knowledge of both its anatomical and functional organisation. For this we need tools and methods that allow us to image for example, gene expression patterns and cell morphology over a broad range of spatial scales. Recent developments in sample preparation and microscopy have opened the door for high-resolution whole brain imaging in small mammals such as mice. This presents new challenges to process and understand an ever increasing deluge of data. This review focusses on recent efforts to analyse and understand mesoscale whole-brain microscopy data with the view to establishing best practice approaches across the imaging community.

### 1.1 Traditional methods

Microscopic imaging of tissue sections has been a key method in neuroscience since the days of Golgi and Cajal. Over the last century, mechanisms for enhancing contrast of brain regions and cell types have been improved, particularly by immunofluorescence (Coons et al., 1941) and the discovery of fluorescent proteins such as GFP (Chalfie et al., 1994). It is now possible, using fluorescence, to distinguish dozens of cell types, and many cellular components. Fluorescence microscopy has also developed quickly and many methods such as traditional wide-field, confocal (Minsky, 1961), multi-photon (Denk et al., 1990) and super-resolution (e.g. Betzig et al., 2006; Rust et al., 2006) techniques are now in routine use in neuroscience laboratories. The limitation of all of these methods for neuroanatomy is that they can only be used to image a relatively thin tissue section, up to around 100µm with confocal microscopy, or up about a millimetre in optimal conditions for multi-photon microscopy (Kobat et al., 2011; Theer et al., 2003). With, for example, single neurons projecting to many regions across the brain (Winnubst et al., 2019) studying small areas can prevent an understanding of the global organisation.

Imaging of large intact brains has been possible for many decades, for example using ultrasound (Donald et al., 1958), computed tomography (Hounsfield, 1973), or magnetic resonance imaging (MRI, Lauterbur, 1973). These traditional three-dimensional imaging methods benefit from being non-invasive, and many aspects of brain activity and its structure can be studied. However, they have two main limitations when it comes to studying detailed organisation. Firstly, they have relatively low spatial resolution which, although continuing to improve, is limited to measurements of gross brain anatomy. Secondly, these methods do not have the specificity required to study all aspects of neuroanatomy using cutting edge viral, genetic and immunofluorescence toolkits.

### 1.2 Whole-brain fluorescence microscopy

To reach a better understanding of neuroanatomy it is necessary to image the entire brain at a sufficient resolution to resolve key structures and with the specificity to distinguish cell types and subcellular features. It is possible to image large volumes of the brain by imaging serial sections, and computationally reconstructing a 3D image volume (e.g. Luzzati et al., 2011). These methods however are laborious, and the manual sectioning process can introduce many artefacts, particularly as the sections can be damaged during processing.

To study the brain's anatomy across spatial scales, high-resolution images acquired from intact brains are required, rather than post-hoc assembly of multiple sections into a single image. There are two broad classes of methods to generate these images (Osten and Margrie, 2013), the first of which are block-face serial sectioning methods. In these methods, an image of the surface of the tissue is taken, and an in-built tissue sectioning system removes a layer of tissue, allowing deeper areas of the brain to be imaged. By acquiring images from the intact brain and then removing a section of tissue to reveal the next part of the brain to be imaged, the disadvantages of traditional sectioning methods can be ameliorated. The sectioning happens after imaging, preventing damage from forming part of the image, and by imaging the intact brain, the individual 2D images are aligned to form a 3D volume without errors introduced by computational reconstruction. Three of the most common methods are STPT (serial two-photon tomography, Ragan et al., 2012, Economo et al., 2016), fMOST (fluorescence micro-optical sectioning tomography, Gong et al., 2013) and FAST (block-face serial microscopy tomography, Seiriki et al., 2017). STPT uses a two-photon microscope to acquire a tiled image of agar-embedded tissue, just below the tissue surface before using a microtome to remove the surface of the tissue, and the process repeats to build up an image of the entire brain. FAST is conceptually similar to STPT, but uses spinning-disk confocal microscopy to increase the speed of data acquisition. In contrast, fMOST uses a diamond knife to remove an ultra-thin section from the surface of a resin embedded brain while a line scan is acquired from the section as it is cut.

The second group of methods used to acquire whole-brain fluorescence microscopy images is the combination of optical tissue clearing and light-sheet fluorescence microscopy (LSFM). Brain tissue can now be optically cleared by the use of organic solvents (Dodt et al., 2007; Ertürk et al., 2012; Renier et al., 2014), lipid removal (Chung et al., 2013; Hama et al., 2015; Susaki et al., 2014) or simple immersion in refractive index matching solutions (Ke et al., 2013; Kuwajima et al., 2013). Rendering the brain optically transparent, along with immunostaining (Chung et al., 2013; S. Kim et al., 2015; Renier et al., 2014), provides a path towards rich high-quality 3D whole-brain datasets. Widefield or traditional point-scanning microscopy is not able to fully exploit the advances in tissue clearing, because of low speeds, and photobleaching due to repeated exposure of the same parts of the tissue. Although based on a very early method (Siedentopf and Zsigmondy, 1903), light sheet imaging has only relatively recently been applied to fluorescence microscopy (Voie et al., 1993). LSFM works by illuminating with a thin sheet of light to excite the fluorophores in an optical section of the tissue. The resulting fluorescence is then detected by a camera positioned orthogonal to the light-sheet. This selective illumination combined with wide-field detection provides speed and reduces photobleaching, allowing repeated rounds of imaging. LSFM has been extensively applied to image whole mouse brains (Dodt et al., 2007; Lerner et al., 2015; Renier et al., 2016, 2014; Susaki et al., 2014; Tomer et al., 2014).

The advances in sample preparation and microscopy over the last decade have now made the acquisition of high-quality whole-brain datasets possible. LSFM and STPT systems are available commercially and through mature open source initiatives (Tomer et al., 2014, Campbell, 2020; Economo et al., 2016; Voigt et al., 2019), ensuring an increasingly large user-base.

Whole-brain microscopy has been simplified but, as many researchers are discovering, many of the challenges begin once the data have been acquired. Tools have been developed to process, visualise, and analyse these new datasets but, as a single brain image can be up to the order of a terabyte, many laboratories remain ill equipped to handle these data. More importantly, traditional open source bioimage analysis tools such as FIJI (Schindelin et al., 2012) and CellProfiler (McQuin et al., 2018) do not have all of the necessary functionality for integrated analyses of all types of whole-brain microscopy data.

2. **Image analysis**

Sample preparation and imaging has been the focus of the field for the last few years, but data analysis is becoming the most difficult challenge. Many neuroscience laboratories do not have extensive image analysis experience, particularly automated analyses of very large datasets. Whole-brain image data sets also have unique challenges, due to the size and specific requirements such as atlas registration. For the purposes of this review, image analysis begins at the point at which a single 3D whole-brain image volume is produced. Image stitching, and artefact correction varies between imaging modalities and is beyond the scope of the review.

Over the past few years many methods have been described in the literature, but very few are designed in a flexible manner and released to the community as user-friendly open-source tools. This review will focus on published tools that are freely available to the community, and which could be easily adopted by a typical laboratory without any high-performance computing resources.

When it comes to analysing whole-brain microscopy datasets, one of the first challenges is segmenting the features of interest. Segmentation refers to the assignment of image voxels to a meaningful label, such as a brain region, a cell, a blood vessel, or any other object. Many of these object segmentation problems have been solved for traditional slice histology, but whole-brain images present new challenges. In particular, the size of the data set, and the subsequent increase in variance in pixel intensities across the brain that can arise from both biological heterogeneity and non-biological fluorescence artefacts.

### 2.1 Neuronal somata

Although there exists a very large number of laboratories focused on the vital function of glia in the maintenance of brain homeostasis, researchers interested in the detection and mapping of neuronal somata have been the main driver for establishing high-throughput imaging pipelines for brain segmentation, cell identification and counting. In addition to mapping the location of neuronal cell types (Mano et al., 2020), such methods are also used for mapping brain activity (Renier et al., 2016) and understanding its cell-to-cell connectivity (Vélez-Fort et al., 2014). Until recently, neuronal cell detection has been performed manually in whole-brain images (Ogawa et al., 2014; Vélez-Fort et al., 2014; Watabe-Uchida et al., 2012), but this method does not scale for routine use, when many thousands of cells can be labelled in each brain. Additionally, manual analyses of this scale are difficult to reproduce, and can introduce an additional source of non-biological variability.

Although conceptually simple, detection of cell bodies in whole-brain images is a complex problem, firstly because the structure to be detected can vary greatly between experiments and cells. For example, label type (nuclear or cytoplasmic), cell size and shape, and the image quality and signal intensity can differ between samples and experiments. There have been two classes of approaches to detect cells in whole-brain datasets. The first is using traditional computer vision approaches such as spatial filters and intensity thresholding. These have been applied in 2D in the WholeBrain (Furth et al., 2018) or AMaSiNe (Song et al., 2020) packages and in 3D in ClearMap (Renier et al., 2016), MIRACL (Goubran et al., 2019) and MagellanMapper (Young et al., 2020), but these methods do not always work well with densely labelled cells or in noisy data. The second class are machine learning approaches. Many studies have used random forest classifiers, implemented using Ilastik (Berg et al., 2019) which has been used in CUBIC-Cloud (Mano et al., 2020) and also in ClearMap. More recently, deep learning (Lecun et al., 2015), and in particular convolutional neural networks (CNNs) have been applied for high-performance cell detection (Iqbal et al., 2019b). These machine-learning approaches however can be slow, and require time-consuming annotation of training data into cell, and non-cell voxels. A recently released method (cellfinder) has combined traditional computer vision approaches for speed, with a deep-learning network to curate the results (Tyson et al., 2020b). In many cases, detecting the position of the cell is all that is required (rather than defining the cell boundaries), and this can be used to generate training data over much shorter timescales (Frasconi et al., 2014).

Cell detection (along with registration, see **section 3**) is an area within whole-brain image analysis with a lot of promising developments though as yet there is no single method that has been shown to work well across all image modalities and label types, and so researchers must trial multiple methods. There are also no methods that allow for identification of cell types (this must be inferred from the input data). In the future, cell detection methods which involve classification (e.g. the machine learning-based methods) could be extended to classify multiple cell types based on morphology, location, and signal intensities.

### 2.2 Neuronal morphology

Sparse labelling of neurons allows for the segmentation and analysis of the morphology of entire cells, including axons and dendrites. Whole-brain datasets should also allow multiple cells to be segmented in their entirety. There are currently no fully-automated methods for neuronal reconstruction in whole-brain microscopy images. Neurons can be traced either manually (Han et al., 2018) or semi-automatically by selecting points along a neurite (Arshadi et al., 2020) or manual connection of algorithmically segmented neurite components (Winnubst et al., 2019). There exist more automated methods (Hang et al. 2018) but these still require human supervision. All of these methods can be very time consuming, and relative to the microscopy, represent a processing bottleneck. Fully automated neuronal reconstruction remains an open challenge.

### 2.3 Connectivity mapping

While many studies investigating connectivity use cell soma detection (e.g. Vélez-Fort et al., 2014, Menegas et al., 2015) or single neuron reconstruction (Winnubst et al., 2019), sometimes the analysis of dense axonal projections is required. The majority of axonal segmentation algorithms only analyse the brain in 2D sections (e.g. The Allen Mouse Brain Connectivity Atlas: Kuan et al., 2015; Oh et al., 2014). To our knowledge, there are two software packages capable of 3D analysis of whole-brain axonal projections. The first is part of the MIRACL toolbox (Goubran et al., 2019) which uses structure tensor analysis to generate streamlines, estimating the diameter of axon bundles. These streamlines can be traced to determine whether they pass through, or terminate within a brain region, and are then used to map connectivity. The second method is TRAILMAP (Friedmann et al., 2020) which uses a 3D CNN (a modified U-Net: Ronneberger et al., 2015) to segment individual axons from the background, allowing axonal density to be quantified across the brain. This method however does not allow tracing of connectivity from one region to another. There is not yet a method allowing individual axons to be traced in these dense datasets, this is currently only possible with sparse labelling.

### 2.4 Vasculature

In addition to segmentation of cell bodies and projections, analysis of the vasculature is important, particularly in preclinical studies such as the study of Alzheimer's disease (Bennett et al., 2018). There have been two methods released recently for the segmentation and analysis of whole-brain vasculature networks. The first is TubeMap (Kirst et al., 2020) which binarises labelled vessels, uses a CNN to fill the resulting image and then performs skeletonisation to produce a map of vessels throughout the brain. The vessels are classified as arteries or veins based on antibody staining, and a computational graph can be constructed to investigate vessel properties such as branching. The other method is VesSAP (Todorov et al., 2020) which uses a fully CNN-based method to segment the vasculature. Both methods appear to perform well, but as yet no studies have compared their differing approaches.

3. **Brain registration and segmentation**

Detecting and locating large numbers of objects (such as all labelled somata) in a whole brain produces a huge amount of data. The obvious way to distil this information and quantify data from multiple animals is to assign the objects to brain regions. The majority of whole-brain microscopy studies now carry out some kind of image segmentation to identify brain structures, and there have been many approaches to solving this problem. The common feature of most of these methods is that they base the segmentation on an existing reference atlas. An atlas typically consists of a reference image (of a single brain, or preferably an average of many), and an associated annotations image, with a mapping from each voxel to a brain region. The standard microscopy reference atlases are traditionally 2D (Dong, 2008; Franklin and Paxinos, 2008), and based on a single animal. While invaluable for many applications, 3D atlases (i.e. a single, aligned 3D volume) are required for computational analysis of whole-brain images.

### 3.1 Image Registration

Registration is usually a key part of a whole-brain microscopy image analysis workflow, and refers to the spatial mapping of an atlas reference image onto the sample data. This can be used for atlas-based segmentation (see **section 3.1.1**), but the sample can also be mapped onto the atlas image. Transforming the sample onto the atlas allows for data from multiple animals to be analysed and visualised in the same coordinate space (**Fig 1**.) which allows a more direct comparison than visualising data side by side, due to inherent variations in brain structure across animals.

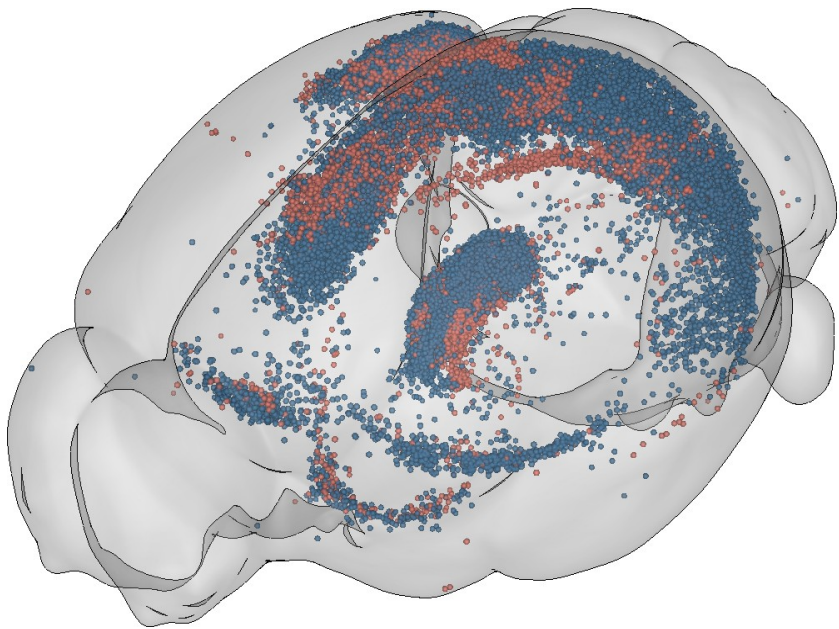

**Fig 1. Warping to atlas space**. Cells detected with cellfinder (Tyson et al. 2020b) from two rabies viral tracing experiments (red and blue), warped to the Allen Mouse Brain Common Coordinate Framework version 3 and visualised using brainrender (Claudi et al. 2020b).

There have been many published pipelines for registration of sample data to an atlas, but the majority have only been used for 2D data. However, there are now software packages released that are suitable for registration of 3D data. Registration packages typically fall into two categories, whether they register the entire image volume to the atlas, or whether they register 2D sections separately. One of the most conceptually simple is the 2D registration method implemented within WholeBrain (Furth et al., 2018). The WholeBrain software detects reference points on a 2D image (at the surface of the brain) and maps these to the surface of the atlas brain. Although this approach works well for 2D data, for 3D whole brain microscopy data this can be time consuming because the user must manually identify the part of the atlas that best matches each image section. To overcome this problem, an extension of WholeBrain has been developed (SMART, Jin et al., 2019) that helps to automate some parts of this manual step. Users can specify the atlas position of the first and last 2D slices in their 3D image and the software can select the atlas planes for the images in between. Although SMART is faster than WholeBrain, the authors estimate that registration of an entire LSFM mouse brain image could still take 3-4 days.

The majority of registration is now carried out using 3D registration tools that are wrappers around existing image registration tools such as NiftyReg (Modat et al., 2010), ANTs (Avants et al., 2011) or Elastix (Klein et al., 2010). These tools typically use a combination of linear (affine) and non-linear (e.g. b-spline) deformations to best match the intensity distributions within the sample and atlas reference images following preprocessing. One of the first methods (ClearMap, Renier et al., 2016) provides a Python interface to Elastix to register LSFM mouse brain images to an atlas (Y. Kim et al., 2015) at a resolution of 25µm. This method has now been updated (Kirst et al., 2020) to use the new Allen Mouse Brain Common Coordinate Framework version 3 (Allen CCFv3, Wang et al., 2020).

Another tool released around the same time is aMAP (Niedworok et al., 2016) which provides a Java interface and FIJI (Schindelin et al., 2012) plugin to the NiftyReg library to register STPT data to a 12.5µm version of the same atlas (Y. Kim et al., 2015). Unlike ClearMap, aMAP was validated against expert manual segmentation. This tool has now been updated, providing a Python interface and a command line tool, along with support for additional atlases (brainreg, Tyson et al., 2020a).

Many more tools have since been developed, such as MIRACL (Goubran et al., 2019) which provides a graphical user interface for ANTs, and has been shown to work well for both LSFM and STPT data (along with other modalities such as MRI). Additionally, MagellanMapper (Young et al., 2020) provides a graphical interface for registration with elastix (via SimpleElastix, Marstal et al., 2016).

There is now work to develop more accurate registration algorithms that do not rely on simple intensity-based approaches, particularly for situations in which the samples are damaged, or the data is contaminated in some other way (e.g. additional signals). These approaches show promise, but as yet their complexity prevents widespread adoption. Tward et al., (2020) developed a pipeline that can infer missing data to best register multi-modal mouse brain image data. There has also been work developing deep learning-based approaches for image registration. Ni et al., (2020) use a CNN to register a sample image to the atlas by combining the mappings of small blocks of the sample image to blocks of the atlas.

### 3.1.1 Segmentation

Segmentation performed on whole-brain microscopy images is typically carried out by registration of an atlas reference image onto the sample image, and then applying the same transform from atlas to sample space to the atlas annotations (**Fig 2.**). The atlas annotations can be overlaid upon the raw image, and used to attribute brain regions. An alternate strategy, first developed for human MRI images is to use CNNs to directly segment the image, without registration to an atlas (Guha Roy et al., 2019; Mehta et al., 2017). There has been one study applying this to microscopy data in mice (albeit traditional 2D data, Iqbal et al., 2019a). This method was used for relatively coarse segmentation of 2D data, but can be used without registration and applied to multiple developmental time points. This method could be applied to 3D microscopy data, and could potentially overcome issues with damaged tissue, or for experiments in which reference atlases do not exist.

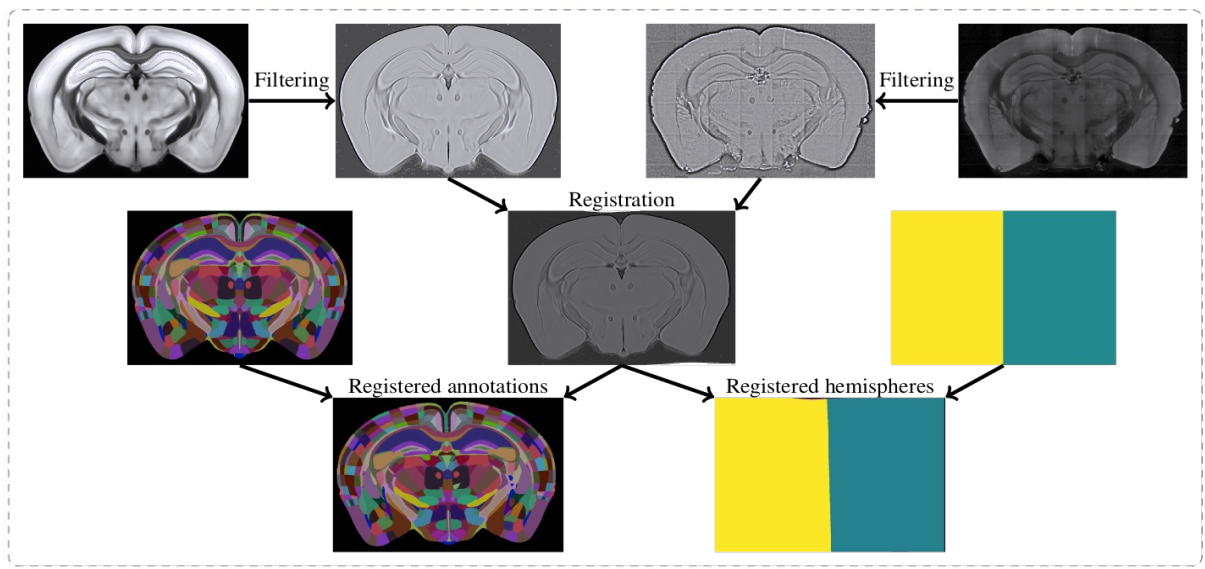

**Fig 2. Atlas-based segmentation.** Atlas reference image (top left) and raw data (top right) are filtered, and the reference image is mapped onto the raw data. Other images, such as the atlas annotations and the brain hemispheres can then be warped similarly onto the raw data.

### 3.2 Reference atlases

3D reference atlases exist for many species, but many of them are not available as a digital 3D image set or are not at a resolution sufficient to take advantage of whole-brain microscopy. Many of them are based on MRI images with relatively low resolution, and some are based on traditional histology with rather modest resolution in the z dimension. For this reason, only high-resolution digital 3D atlases based on microscopy data, or other atlases that have been used for processing of whole-brain microscopy data will be discussed.

The majority of 3D whole-brain microscopy atlases are in mice, and by far the most commonly used is the Allen CCFv3 (**Fig. 3A**). This atlas consists of a reference image (with 10μm isotropic voxels), generated from 1,675 STPT images, and an annotations image, delineating 658 different brain regions (including isocortical areas, subcortical structures, fibre tracts and ventricles) defined by transgenic reporter mice and axonal projection data along with *in situ* hybridisation, antibody staining and traditional cytoarchitectural stains such as Nissl.

The Allen CCFv3 is comprehensive, but is still missing many brain region subdivisions. For this reason, an additional atlas was developed (Chon et al., 2019) in the same coordinate space, but with additional annotations (**Fig. 3B**). Primarily, additional regions were added from the Franklin-Paxinos atlas (Franklin and Paxinos, 2008) along with additional experimental data (e.g. additional transgenic lines) and the striatum was further subdivided based on connectivity data in the literature (Hintiryan et al., 2016; Hunnicutt et al., 2016; Oh et al., 2014). Unlike the Allen atlas, this atlas is not defined at high resolution isotropically, but the additional annotations will be valuable for many studies (e.g. of the striatum).

A different way to define an atlas is by directly using gene expression, rather than a mix of gene expression, protein expression, cytoarchitecture and connectivity. A recent atlas (Ortiz et al., 2020) uses spatial transcriptomics (Ståhl et al., 2016), followed by clustering methods for an unsupervised, data-driven approach to subdividing the brain into meaningful regions (**Fig. 3C**). This atlas is also in the same coordinate space as the Allen CCFv3, but the annotations differ considerably. It remains to be seen whether these delineations fit better with other data (e.g. electrophysiological cell properties) than the traditional methods for atlas generation.

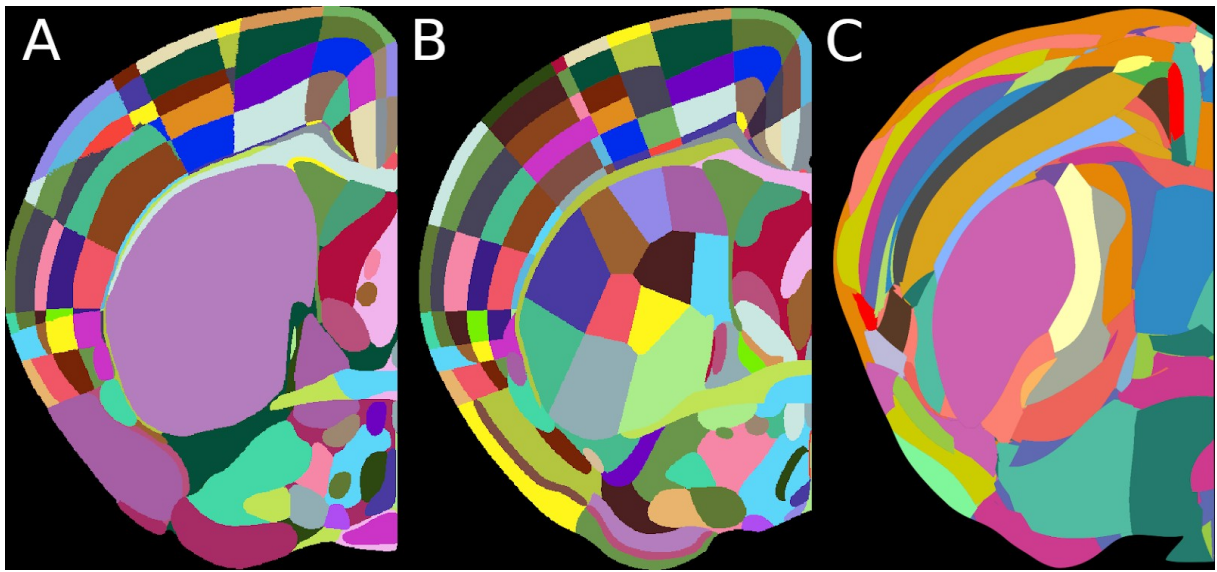

**Fig 3. Comparison of mouse atlases. Single hemisphere section at bregma.** A) Allen Mouse Brain Common Coordinate Framework version 3 (Wang et al., 2020). B) Enhanced and Unified Mouse Brain Atlas (Chon et al., 2019). C) Molecular atlas of the adult mouse brain (Ortiz et al., 2020)

Although the majority of rodent whole-brain imaging is carried out in mice, other model species are beginning to be used, such as the rat (Branch et al., 2019; Stefaniuk et al., 2016). There aren't any rat atlases of the same quality of the mouse atlases, but there are high quality atlases based on MRI (e.g. the the Waxholm Space atlas, Papp et al., 2014). To our knowledge, there has been one study describing a method to register rat LSFM hemisphere images to this atlas (Branch et al., 2019), however this atlas is of relatively low resolution, and only consists of 76 subdivided regions.

Another group of model species that are being used with whole-brain microscopy are monkeys, such as marmosets (Skibbe et al., 2019; Susaki et al., 2014, 2020). There are efforts to create high-resolution marmoset atlases by combining data from MRI and traditional microscopy techniques (Majka et al., 2020; Woodward et al., 2018), but as yet these atlases have not yet been used with LSFM or STPT data.

As the tissue clearing and imaging methods evolve, more and more "novel" species will be imaged. Without atlas development, the insights that can be gained from these species will be limited compared to mice, for which there have been many resources developed.

### 3.2.3 Atlas reference images

All atlases come with a template, reference image upon which the annotations are based. In whole-brain microscopy, this reference image becomes critical, because it is used for registration of sample data into the atlas coordinate space. The reference image of the Allen CCFv3 is a STPT image, and the atlases that are either based on it, or warped to it (Chon et al., 2019; Ortiz et al., 2020) also use the same image. Registration of other STPT images to this template therefore works well (Y. Kim et al., 2015; Niedworok et al., 2016), but other imaging modalities (e.g. LSFM) may not work as well. Data preprocessing may help improve registration performance, but LSFM images of cleared tissue are considerably different to STPT images. One study has addressed this (Perens et al., 2020), by developing a warped version of the Allen CCFv3 with an LSFM template. The template was generated from 139 mouse brains cleared with iDISCO+ (Renier et al., 2014, 2016). To overcome non-uniform morphological changes following clearing, the authors individually registered six brain regions from the Allen CCFv3 template to the LSFM template. The Allen CCFv3 annotations were warped similarly, and so the LSFM atlas can be used directly with LSFM data, without transforming the data into the original Allen CCFv3 coordinate space. As other atlases also use the Allen CCFv3 template image, these annotations (e.g. Chon et al., 2019; Ortiz et al., 2020) could also be warped into the space of the LSFM atlas. Unlike STPT, LSFM data can vary considerably, partially due to the different microscopes available, but mostly due to the tissue clearing method. Different clearing methods rely on different mechanisms to render the tissue transparent (affecting contrast) and can differentially affect brain size (Wan et al., 2018), potentially causing morphological changes around the ventricles. This atlas for iDISCO+ cleared brains is likely useful for iDISCO+ samples, but similar atlases will likely need to be developed for the different families of clearing methods.

### 3.2.4 Alternative atlas formats

The majority of brain atlases rely on reference images, with a corresponding annotations image, defined by a raster image (Perens et al., 2020), polygons (Wang et al., 2020) or smooth curves (Furth et al., 2018). As the resolution of whole-brain microscopy data increases, so will the file sizes of the atlases required, along with the computational requirements for data processing. There are alternative strategies, such as by defining an atlas based on the coordinates of every cell in the brain (Murakami et al., 2018), rather than an image of the brain. The authors of this atlas used expansion microscopy (Chen et al., 2015) to generate a very-high resolution image of the mouse brain, and then segmented every cell in the image. The 14TB image can be represented by point clouds, taking up less than 3GB, however data must be prepared in a particular way to use atlases of this form (an image must exist with all cells labelled). It remains to be seen whether this approach will become as widespread as image-based atlases, but tools are being developed to take advantage of this approach (Mano et al., 2020).

### 3.2.5 Choosing an atlas

Traditionally, atlases suitable for whole-brain microscopy were rare, and there wasn't much choice available. This is gradually changing, and so users will need to choose the most appropriate atlas for their work. In some cases this is obvious, the atlas must be for the model species being imaged, but in other cases, a choice must be made. Existing atlases have been developed in relative isolation, and as such they are organised in different ways, use different file types, and in most cases are not interoperable. This makes it difficult for researchers to choose the most appropriate atlas, as their analysis pipelines must be rewritten to make use of a new atlas. The MRI community has been using atlases in this way for much longer, and so there is ongoing work to standardise atlases and make them available (Bakker et al., 2015; Myers et al., 2019). More recently, efforts towards standardising analysis and atlas usage have developed for the whole-brain microscopy field, including the natverse (Bates et al., 2020) and BrainGlobe (Claudi et al., 2020a) projects. The BrainGlobe project provides a Python application programming interface that provides a number of atlases in a standard format, allowing users to switch between them. Work to further standardise the generation and release of atlases will be required to simplify their use, and allow the correct atlas to be chosen.

## 4. Data visualisation

### 4.1 Raw data

Whole-brain microscopy presents new challenges for data visualisation. The first challenge is visualisation of the raw data. Compared to traditional microscopy methods, this is far more difficult as the majority of images do not fit in the memory of most computers. Luckily this is a challenge faced by many other imaging fields, and so there are existing strategies to handle this data. The simplest way is to use so-called "lazy loading" of 2D image data. Software such as FIJI (Schindelin et al., 2012) and napari (Sofroniew et al., 2020) allow users to scroll through large 3D images plane by plane, and only the 2D section being viewed at a time is loaded into memory. This approach is useful for visualising data quality, but it can be slow and does not provide any 3D information. An alternative strategy is to use alternative file formats that store "chunks" of 3D data at different resolutions. This allows a 3D low-resolution overview to be viewed, and only the data in the field of view is required to be loaded into memory when the user zooms in. This approach is implemented in many commercial software packages and the open-source BigDataViewer (Pietzsch et al., 2015) plugin for FIJI.

### 4.2 Segmented data

One of the biggest challenges is specific to whole-brain microscopy: 3D visualisation of segmented data in a common coordinate space. These datasets are very complex, potentially containing segmented cells, neurites, brain regions and implanted devices (e.g. Neuropixels probes, Jun et al., 2017). Registration to a common atlas space allows for data from multiple samples to be viewed together (see **Fig. 1**) further complicating the data to be visualised. Often these data cannot be easily understood in 2D, and so 3D tools which allow visualisation of arbitrary shapes within an atlas coordinate system are required.

Many of the existing packages for whole-brain microscopy analysis include some tools for visualising segmented data along with an atlas (Furth et al., 2018; Kirst et al., 2020) but these are often limited to the data analysed within the software itself and require some programming knowledge. There are also packages released for visualising data from specific atlases, such as the Allen CCFv3 (https://connectivity.brain-map.org/3d-viewer, https://github.com/AllenInstitute/cocoframer, https://github.com/Yaoyao-Hao/BrainMesh), but these cannot be used with other atlases, and are limited to what additional data (other than brain structures) can be visualised.

More recently, tools have been developed that allow integration of both publicly available datasets (such as the MouseLight project, Winnubst et al., 2019) along with user-generated data. The natverse (Bates et al., 2020) provides functionality for analysis and visualisation of neuronal morphology, although many of the functions are specific to Drosophila. The SNT FIJI toolbox (Arshadi et al., 2020) allows analysis of neuronal morphology and visualisation of atlas structures and reconstructed neurons from multiple projects in Drosophila, zebrafish and mouse. Lastly, brainrender (Claudi et al., 2020b) provides functionality to visualise publicly available data, atlas data and user data, using the same code to visualise data across species. Brainrender is part of the BrainGlobe project (Claudi et al., 2020a) to support multiple atlases, and integration with other software such as brainreg (Tyson et al., 2020a) and cellfinder (Tyson et al., 2020b).

## 5. Outstanding needs

### 5.1 Additional analyses

Whole-brain microscopy is becoming more common, and is being applied more broadly, but user-friendly tools are not available for all types of analyses. While there are many tools available for registration and segmentation of common structures such as neuronal somata, they do not exist for other structures or classes of cells. Structures of a similar size to neuronal somata, such as amyloid plaques may be detected with existing cell-detection algorithms (Liebmann et al., 2016), but other types of structure cannot. More complex structures such as glial cells are difficult to segment, and there is a need for dedicated tools so that imaging advances can be used to study glia in the same way as neuronal cells. Existing segmentation algorithms are also designed to detect a single type of structure from a single image channel, and mostly cannot distinguish different structures within a single image, although some tools can detect both cell somata and dendrites (e.g. Furth et al., 2018).

Segmentation of large structures (such as lesions, injection sites and implanted devices) is computationally straightforward, but existing software packages do not include such methods and so users must create their own pipelines. The introduction of more complex implanted devices such as Neuropixels probes (Jun et al., 2017) with hundreds of closely-packed recording sites, necessitates the precise mapping of such objects within a common coordinate space from whole-brain microscopy data (Liu et al., 2020). Manual interrogation is possible within the BrainGlobe suite (Tyson et al., 2020a), but general purpose, automated mapping of these devices onto the segmented brain is not yet available.

Most of the existing analysis pipelines for whole-brain microscopy are conceptually simple, their advancement is to be able to deal with the scale and heterogeneity of the data. However, much more sophisticated analyses are plausible. Rather than simple cell detection, some algorithms could be adapted to classify cell types based on morphology. This would allow for much richer information to be extracted from these datasets without antibody staining.

### 5.2 Tool comparison

Many laboratories are now faced with a large amount of data, and a confusing landscape of analysis tools to choose from. While in some cases (e.g. vessel segmentation) the number of tools available are relatively limited, in other areas (e.g. registration) there are many tools, with no obvious answers as to which method is the most suited for a particular application. Comparisons between tools exist in the literature, but these may be biased as they are carried out by the developers of a single tool. It is difficult for a single researcher or team to produce an objective comparison of different analysis tools, so it is good practice to invite tool developers to "compete" to produce the best results on a set of benchmark data (e.g. Sage et al., 2019; Ulman et al., 2017). If the original developers of the software carry out the analysis, they are incentivised to produce the best results, and the interested user can see the theoretical best performance of each tool on standardised data. This has not yet been carried out for any aspect of whole-brain microscopy analysis, but will likely be necessary as many more tools are developed. In some cases, generating a metric of accuracy for validation purposes is relatively simple (e.g. cell counting), but in others (e.g. brain region segmentation) it can be much more difficult (Niedworok et al., 2016).

### 5.2.1 Neuronal somata detection

Neuronal somata detection is one of the most common whole-brain microscopy image analysis tasks, but each tool was originally developed for different types of data, such as nuclear cfos activity mapping (Renier et al., 2016) or whole cell labelling in viral tracing experiments (Tyson et al., 2020b). Unless the user has very similar data to that which is described in the software's publication or documentation, it is not clear even which packages should be tested. This will gradually become clearer as more studies are published using these tools, but until then it remains difficult to compare the performance of multiple algorithms.

### 5.2.2 Registration and segmentation

Many of the existing methods quantify registration performance (Goubran et al., 2019; Iqbal et al., 2019a; Ni et al., 2020; Niedworok et al., 2016) or compare to other tools (Goubran et al., 2019; Iqbal et al., 2019a; Ni et al., 2020), but these are limited in their utility for the user who is choosing which software to use. It is common in MRI registration to compare algorithms across many datasets (Klein et al., 2009), but this is much more complex for whole-brain microscopy. The main reason is that the community has not yet decided on measures to assess registration and segmentation accuracy. Many measures have been used such as comparison to expert region segmentation and landmark registration error (Niedworok et al., 2016, Goubran et al., 2019). Until a standardised set of measures is defined which captures all aspects of registration accuracy, users must use trial and error to find the most appropriate tool.

### 5.3 Workflow integration

There are many tools available for the analysis of whole-brain microscopy data (**Table 1**), and some packages can be used for multiple types of analysis (e.g. Goubran et al., 2019), but none of these provide an integrated workflow for all types of analyses. Unlike many other types of microscopy, whole-brain microscopy images may contain different features across spatial scales that need to be segmented and analysed. A single image could contain injection sites and labelled cells along with lesions and implanted devices. To fully analyse the data, all of these features must be segmented and analysed in a common coordinate space. In contrast to traditional image analysis packages in which all necessary analyses can often be carried out (McQuin et al., 2018; Schindelin et al., 2012), whole-brain image analysis must be carried out with multiple packages and combined by the user. This process is time-consuming and technically difficult because it relies on custom pipelines to be developed by each laboratory. Such pipelines are rarely re-used by the community.

The majority of software in this field is developed by academics, for whom publishing a paper is often the most important end result. There is rarely funding for continued software development and refinement, and so the software is often more difficult to use than necessary, does not interface with other software packages, and does not always use the most up to date technologies. To overcome these issues, without requiring an onerous amount of work, we propose three well-established techniques for increasing interoperability, and reducing duplication of effort. These are common file formats, software packages and plugin systems.

Increasing interoperability of software packages will have two main advantages for the community. The first is that users can combine different types of analysis within a single workflow (e.g. cell detection and vessel segmentation). The second is that it will allow direct comparison of different approaches to the same problem. In the case of cell detection, there are many different methods, each of which was developed for different types of data and cellular markers. It is likely that one of these will be the most successful for an individual dataset, but it is time consuming to directly compare methods on a single dataset.

An increase in interoperability will make it easier for users to compare algorithms (e.g. by visualising results in the same software), and create an integrated pipeline by selecting the most appropriate parts of existing software packages.

| Software package | Reference | Website | Implementation | Registration | Supported atlases | Cell detection | Axon tracing | Vasculature segmentation | Visualisation |
|---|---|---|---|---|---|---|---|---|---|
| ClearMap/ ClearMap2 | Renier et al. 2016 Kirst et al. 2020 | christophkirst.github.io/ ClearMap2Documentation | Python | 3D using Elastix | Allen Mouse Brain (25um) | 3D – nuclei | N/A | Vessel segmentation & analysis | In-built tools |
| WholeBrain | Furth et al. 2018 | wholebrainsoftware.org | R | 2D using reference points | Custom, based on Allen Mouse Brain | 2.5D – whole cell | N/A | N/A | In-built tools |
| MIRACL | Goubran et al. 2019 | miracl.readthedocs.io | Python | 3D using ANTs | Allen Mouse Brain (25um) | 3D nuclei & whole cell | Bulk streamline analysis | N/A | In-built tools |
| AMaSiNe | Song et al. 2020 | github.com/vsnnlab/ AMaSiNe | MATLAB | 2D using Elastix | Allen Mouse Brain (25um) | 2D – nuclei | N/A | N/A | In-built tools |
| cellfinder | Tyson et al. 2020 | cellfinder.info | Python | 3D using brainreg | Multiple, via BrainGlobe | 3D – whole cell | N/A | N/A | Export to napari & brainrender |
| TRAILMAP | Friedmann et al. 2020 | github.com/AlbertPun/ TRAILMAP | Python | N/A | N/A | N/A | Axon segmentation | N/A | N/A |
| VesSAP | Todorov et al. 2020 | github.com/vessap/ vessap | Python | N/A | N/A | N/A | N/A | Vessel segmentation & analysis | N/A |
| Magellan Mapper | Young et al. 2020 | github.com/sanderslab/ magellanmapper | Python | 3D using SimpleElastix | Multiple | 3D – nuclei | N/A | N/A | In-built tools |
| SNT | Arshadi et al. 2020 | imagej.net/SNT | Java (FIJI plugin) | N/A | N/A | N/A | Single cell tracing | N/A | In-built tools |

**Table 1. Comparison of selected whole-brain microscopy analysis tools**

### 5.3.1 Common file formats

One of the easiest ways to increase interoperability of different software packages is by the use of common file formats. Although many packages carry out the same type of analyses, the data is stored in different ways, and as such it can be challenging to visualise. In some cases, the actual file type is different (e.g. NifTI vs TIFF for storing registration results), but in other cases the format of the underlying data also changes (e.g. the image origin for cell somata coordinates). Converting formats often requires programming knowledge and for the user to spend time understanding the underlying format. Deciding upon common formats (even as optional exports from the software) would immediately allow analysis using multiple packages, and visualisation and comparison in a single visualisation environment. For most aspects of whole-brain image analysis, this would be relatively simple, as the majority of files saved are 3D images, points or surfaces, for which existing standards are available.

### 5.3.2 Common software packages

Although each new software package contains novel analysis algorithms, much of the code is repeated from one tool to another. Routines such as loading and saving data and assigning detected features to an atlas are common to nearly all software. If these tools were centralised and available for use by the community in isolation from specific analysis packages, developers could save lots of time rather than reinventing the wheel. A useful side effect would be that by adopting these common solutions, new software would naturally become more interoperable, as they are written to be compatible with the same common software packages. The use of common software packages is standard practice in all areas of computer science, including microscopy analysis, such as ImgLib2 for FIJI (Pietzsch et al., 2012) and scikit-image in Python (Van Der Walt et al., 2014).

The only existing package specifically for whole-brain microscopy is the BrainGlobe Atlas API (Claudi et al., 2020a), providing a common Python-based interface for downloading, managing and interfacing with neuroanatomical atlases. Software using this package (Claudi et al., 2020b; Tyson et al., 2020a; Tyson et al., 2020b) can simply reuse code for using atlases and for defining neuroanatomical conventions. If packages such as this were widely adopted by the community, it would reduce the burden of developing new software, and increase operability.

### 5.3.3 Plugin systems

A logical extension of using separate software packages for common tasks, is to develop plugins for existing software. This provides the benefits of developing a central, community-managed software package for all whole-brain microscopy analysis tasks, without the prohibitive amount of effort that would be involved in coordinating such an effort. The plugin ecosystem has been very successful for FIJI, and some whole-brain analysis packages have been written as FIJI plugins (Arshadi et al., 2020; Niedworok et al., 2016), but most recent packages are written in Python (Kirst et al., 2020; Renier et al., 2016; Todorov et al., 2020; Tyson et al., 2020b; Young et al., 2020). Although it is possible to create FIJI plugins based on Python code, reducing the amount of effort to develop compatible software will be key to increasing interoperability.

Napari (Sofroniew et al., 2020) is a new Python-based image viewer, created with the visualisation and analysis of large microscopy images in mind. One of the aims of napari is to develop a plugin architecture to leverage the growing community of image analysis packages developed in Python and provide a user friendly graphical user interface and interoperability between software. Adopting existing software like napari, in which many difficult problems have been solved (such as visualisation of large multichannel images) would also reduce time taken to develop new packages and would increase the potential for interoperability between software. More importantly, a user friendly interface would encourage users to adopt new methods and exploit the benefits of recent developments in sample preparation and imaging.

### 5.4 Communication and collaboration

A typical whole-brain microscopy study may generate many TB of data from multiple brains. For a cell detection workflow, this can be distilled down to cell positions in atlas space, but this still presents an issue when communicating the results to collaborators or in a publication. Simple 2D figures are limited, and it is difficult to fully comprehend text summaries of cellular distributions. 3D, interactive, explorable and shareable data summaries are required to allow others to fully appreciate the data. Interactive web visualisations can be exported in 2D using WholeBrain or 3D using brainrender, but these are not fully customisable, and do not yet allow the user to explore arbitrary properties of the data.

The whole-brain microscopy field has shared protocols for tissue preparation and imaging, along with data analysis tools. However, the majority of these data analysis tools rely on complex machine learning algorithms which require training data. Many existing machine learning tools provide repositories for sharing training data and trained models (e.g. http://www.mousemotorlab.org/dlc-modelzoo, https://bioimage.io). A community effort to host and share such data for whole-brain microscopy segmentation methods would further reduce barriers to entry for those new to the field.

### 6. Conclusions

The field of whole-brain microscopy in small mammals has exploded in the last ten years following advances in tissue clearing and microscopy, although analysis tools have lagged somewhat. There have been many tools for registration and segmentation, but only a limited few such as WholeBrain (Furth et al., 2018) and MIRACL (Goubran et al., 2019) are user-friendly enough to be widely adopted by neuroscientists. These datasets may contain huge amounts of information (labelled cells, neurites, vasculature, implanted devices etc.), but there is no single platform that allows a user to perform all of these analyses. Although some tools have shown promise for integrating multiple types of analysis (Goubran et al., 2019; Kirst et al., 2020; Tyson et al., 2020), there is not yet a platform that allows for them to be combined along with other custom analyses. We propose that this problem can be solved with collaboration and the development of open standards and plugins for existing software.

### Acknowledgements

This work was supported by grants from the Gatsby Charitable Foundation (GAT3361) and Wellcome Trust (090843/F/09/Z and 214333/Z/18/Z) to T.W.M. We thank Rob Campbell for his valuable feedback on the manuscript.

**References**

Arshadi, C., Günther, U., Eddison, M., Harrington, K., Ferreira, T., 2020. SNT: A Unifying Toolbox for Quantification of Neuronal Anatomy 1–16. https://doi.org/10.1101/2020.07.13.179325

Avants, B.B., Tustison, N.J., Song, G., Cook, P.A., Klein, A., Gee, J.C., 2011. A reproducible evaluation of ANTs similarity metric performance in brain image registration. Neuroimage 54, 2033–2044. https://doi.org/10.1016/j.neuroimage.2010.09.025

Bakker, R., Tiesinga, P., Kötter, R., 2015. The Scalable Brain Atlas: Instant Web-Based Access to Public Brain Atlases and Related Content. Neuroinformatics 13, 353–366. https://doi.org/10.1007/s12021-014-9258-x

Bates, A.S., Manton, J.D., Jagannathan, S.R., Costa, M., Schlegel, P., Rohlfing, T., Jefferis, G.S.X.E., 2020. The natverse, a versatile toolbox for combining and analysing neuroanatomical data. Elife 9, 1–35. https://doi.org/10.7554/eLife.53350

Bennett, R.E., Robbins, A.B., Hu, M., Cao, X., Betensky, R.A., Clark, T., Das, S., Hyman, B.T., 2018. Tau induces blood vessel abnormalities and angiogenesis-related gene expression in P301L transgenic mice and human Alzheimer's disease. Proc. Natl. Acad. Sci. U. S. A. 115, E1289–E1298. https://doi.org/10.1073/pnas.1710329115

Berg, S., Kutra, D., Kroeger, T., Straehle, C.N., Kausler, B.X., Haubold, C., Schiegg, M., Ales, J., Beier, T., Rudy, M., Eren, K., Cervantes, J.I., Xu, B., Beuttenmueller, F., Wolny, A., Zhang, C., Koethe, U., Hamprecht, F.A., Kreshuk, A., 2019. Ilastik: Interactive Machine Learning for (Bio)Image Analysis. Nat. Methods. https://doi.org/10.1038/s41592-019-0582-9

Betzig, E., Patterson, G.H., Sougrat, R., Lindwasser, O.W., Olenych, S., Bonifacino, J.S., Davidson, M.W., Lippincott-Schwartz, J., Hess, H.F., 2006. Imaging intracellular fluorescent proteins at nanometer resolution. Science (80-. ). 313, 1642–1645. https://doi.org/10.1126/science.1127344

Branch, A., Tward, D., Vogelstein, J.T., Wu, Z., Gallagher, M., 2019. An optimized protocol for iDISCO + rat brain clearing , imaging , and analysis.

Campbell, R.A.A., 2020. BakingTray: Serial-section automated anatomy extension for ScanImage. https://doi.org/10.5281/zenodo.3631610

Chalfie, M., Tu, Y., Euskirchen, G., Ward, W.W., Prasher, D.C., 1994. Green fluorescent protein as a marker for gene expression. Science (80-. ). 263, 802–805. https://doi.org/10.1126/science.8303295

Chen, F., Tillberg, P.W., Boyden, E.S., 2015. Expansion microscopy. Science (80-. ). 347.

Chon, U., Vanselow, D.J., Cheng, K.C., Kim, Y., 2019. Enhanced and Unified Anatomical Labeling for a Common Mouse Brain Atlas. Nat. Commun. 5067. https://doi.org/10.1101/636175

Chung, K., Wallace, J., Kim, S., Kalyanasundaram, S., Andalman, A.S., Davidson, T.J., Mirzabekov, J.J., Zalocusky, K.A., Mattis, J., Denisin, A.K., Pak, S., Bernstein, H., Ramakrishnan, C., Grosenick, L., Gradinaru, V., Deisseroth, K., 2013. Structural and molecular interrogation of intact biological systems. Nature 497, 332–337. https://doi.org/10.1038/nature12107

Claudi, F., Petrucco, L., Tyson, A. L., Branco, T., Margrie, T.W., Portugues, R., 2020a. BrainGlobe Atlas API: a common interface for neuroanatomical atlases. J. Open Source Softw. 5, 2668. https://doi.org/10.21105/joss.02668

Claudi, F., Tyson, A.L., Petrucco, L., Margrie, T.W., Portugues, R., Branco, T., 2020b. *Brainrender*: a python-based software for visualizing anatomically registered data. bioRxiv. https://doi.org/10.1101/2020.02.23.961748

Coons, A., Creech, H.J., Jones, R., 1941. Immunological properties of an antibody containing a fluorescent group. Proc. Soc. Exp. Biol. Med. 47, 200–202.

Denk, W., Strickler, J.H., Webb, W.W., 1990. Two-Photon Laser Scanning Fluorescence Microscopy. Science (80-. ). 248, 73–76.

Dodt, H., Leischner, U., Schierloh, A., 2007. Ultramicroscopy: three-dimensional visualization of neuronal networks in the whole mouse brain. Nat. Methods 4, 331–336. https://doi.org/10.1038/NMETH1036

Donald, I., Macvicar, J., Brown, T.G., 1958. Investigation of abdominal masses by pulsed ultrasound. Lancet 1, 1188–1195. https://doi.org/10.1016/S0140-6736(58)91905-6

Dong, H.W., 2008. The Allen reference atlas: A digital color brain atlas of the C57Bl/6J male mouse., The Allen reference atlas: A digital color brain atlas of the C57Bl/6J male mouse. John Wiley & Sons Inc, Hoboken, NJ, US.

Economo, M.N., Clack, N.G., Lavis, L.D., Gerfen, C.R., Svoboda, K., Myers, E.W., Chandrashekar, J., 2016. A platform for brain-wide imaging and reconstruction of individual neurons. Elife 5, 2015–2017. https://doi.org/10.7554/eLife.10566

Ertürk, A., Becker, K., Jährling, N., Mauch, C.P., Hojer, C.D., Egen, J.G., Hellal, F., Bradke, F., Sheng, M., Dodt, H.-U., 2012. Three-dimensional imaging of solvent-cleared organs using 3DISCO. Nat. Protoc. 7, 1983–1995. https://doi.org/10.1038/nprot.2012.119

Franklin, K.B.J., Paxinos, G., 2008. The Mouse Brain in Stereotaxic Coordinates, 3rd. ed.

Frasconi, P., Silvestri, L., Soda, P., Cortini, R., Pavone, F.S., Iannello, G., 2014. Large-scale automated identification of mouse brain cells in confocal light sheet microscopy images. Bioinformatics 30, i587–i593. https://doi.org/10.1093/bioinformatics/btu469

Friedmann, D., Pun, A., Adams, E.L., Lui, J.H., Kebschull, J.M., Grutzner, S.M., Castagnola, C., Tessier-Lavigne, M., Luo, L., 2020. Mapping mesoscale axonal projections in the mouse brain using a 3D convolutional network. Proc. Natl. Acad. Sci. U. S. A. 117, 11038–11047. https://doi.org/10.1073/pnas.1918465117

Furth, D., Vaissiere, T., Tzortzi, O., Xuan, Y., Martin, A., Lazaridis, I., Spigalon, G., Fisone, Gi., Tomer, R., Deisseroth, K., Carlen, M., Miller, C.A., Rumbaugh, G., Meletis, K., 2018. An interactive framework for whole-brain maps at cellular resolution. Nat. Neurosci. 21, 139–149. https://doi.org/10.2514/6.2011-3838

Gong, H., Zeng, S., Yan, C., Lv, X., Yang, Z., Xu, T., Feng, Z., Ding, W., Qi, X., Li, A., Wu, J., Luo, Q., 2013. Continuously tracing brain-wide long-distance axonal projections in mice at a one-micron voxel resolution. Neuroimage 74, 87–98. https://doi.org/10.1016/j.neuroimage.2013.02.005

Goubran, M., Leuze, C., Hsueh, B., Aswendt, M., Ye, L., Tian, Q., Cheng, M.Y., Crow, A., Steinberg, G.K., McNab, J.A., Deisseroth, K., Zeineh, M., 2019. Multimodal image registration and connectivity analysis for integration of connectomic data from microscopy to MRI. Nat. Commun. 10, 1–17. https://doi.org/10.1038/s41467-019-13374-0

Guha Roy, A., Conjeti, S., Navab, N., Wachinger, C., 2019. QuickNAT: A fully convolutional network for quick and accurate segmentation of neuroanatomy. Neuroimage 186, 713–727. https://doi.org/10.1016/j.neuroimage.2018.11.042

Hama, H., Hioki, H., Namiki, K., Hoshida, T., Kurokawa, H., Ishidate, F., Kaneko, T., Akagi, T., Saito, T., Saido, T., Miyawaki, A., 2015. ScaleS: an optical clearing palette for biological imaging. Nat. Neurosci. 18, 1518–1529. https://doi.org/10.1038/nn.4107

Han, Y., Kebschull, J.M., Campbell, R.A.A., Cowan, D., Imhof, F., Zador, A.M., Mrsic-Flogel, T.D., 2018. The logic of single-cell projections from visual cortex. Nature 556, 51–56. https://doi.org/10.1038/nature26159

Hang, Z., Anan, L., Feng, X., Ning, L., Jiacheng, H., Hongtao, K., Yijin, C., Yun, L., Wenqian, F., Yidong, L., Huimin, L., Sen, J., Zhiming, L., Fuqiang, X., Yu-hui, Z., Xiaohua, L., Xiuli, L., Hui, G., Qingming, L., Tingwei, Q., Shaoqun, Z. 2018 bioRxiv; doi: https://doi.org/10.1101/223834

Hintiryan, H., Foster, N.N., Bowman, I., Bay, M., Song, M.Y., Gou, L., Yamashita, S., Bienkowski, M.S., Zingg, B., Zhu, M., Yang, X.W., Shih, J.C., Toga, A.W., Dong, H.-W., 2016. The mouse cortico-striatal projectome. Nat. Neurosci. 19, 1100–1114.

Hounsfield, G.N., 1973. Computerized transverse axial scanning (tomography): I. Description of system. Br. J. Radiol. 46, 1016–1022. https://doi.org/10.1259/0007-1285-46-552-1016

Hunnicutt, B.J., Jongbloets, B.C., Birdsong, W.T., Gertz, K.J., Zhong, H., Mao, T., 2016. A comprehensive excitatory input map of the striatum reveals novel functional organization. Elife 5, 1–32. https://doi.org/10.7554/eLife.19103

Iqbal, A., Khan, R., Karayannis, T., 2019a. Developing a brain atlas through deep learning. Nat. Mach. Intell. 1, 277–287. https://doi.org/10.1038/s42256-019-0058-8

Iqbal, A., Sheikh, A., Karayannis, T., 2019b. DeNeRD: high-throughput detection of neurons for brain-wide analysis with deep learning. Sci. Rep. 9, 1–13. https://doi.org/10.1038/s41598-019-50137-9

Jin, M., Nguyen, J.D., Weber, S.J., Mejias-Aponte, C.A., Madangopal, R., Golden, S.A., 2019. SMART: An open source extension of WholeBrain for iDISCO+ LSFM intact mouse brain registration and segmentation. bioRxiv. https://doi.org/10.1101/727529

Jun, J.J., Steinmetz, N.A., Siegle, J.H., Denman, D.J., Bauza, M., Barbarits, B., Lee, A.K., Anastassiou, C.A., Andrei, A., Aydin, Ç., Barbic, M., Blanche, T.J., Bonin, V., Couto, J., Dutta, B., Gratiy, S.L., Gutnisky, D.A., Häusser, M., Karsh, B., Ledochowitsch, P., Lopez, C.M., Mitelut, C., Musa, S., Okun, M., Pachitariu, M., Putzeys, J., Rich, P.D., Rossant, C., Sun, W.L., Svoboda, K., Carandini, M., Harris, K.D., Koch, C., O'Keefe, J., Harris, T.D., 2017. Fully integrated silicon probes for high-density recording of neural activity. Nature 551, 232–236. https://doi.org/10.1038/nature24636

Ke, M., Fujimoto, S., Imai, T., 2013. SeeDB: a simple and morphology-preserving optical clearing agent for neuronal circuit reconstruction. Nat. Neurosci. 16, 1154–1161. https://doi.org/10.1038/nn.3447

Kim, S., Cho, J.H., Murray, E., Bakh, N., Choi, H., Ohn, K., Ruelas, L., Hubbert, A., McCue, M., Vassallo, S.L., Keller, P.J., Chung, K., 2015. Stochastic electrotransport selectively enhances the transport of highly electromobile molecules. Proc. Natl. Acad. Sci. E6274–E6283. https://doi.org/10.1073/pnas.1510133112

Kim, Y., Venkataraju, K.U., Pradhan, K., Mende, C., Taranda, J., Turaga, S.C., Arganda-Carreras, I., Ng, L., Hawrylycz, M.J., Rockland, K.S., Seung, H.S., Osten, P., 2015. Mapping social behavior-induced brain activation at cellular resolution in the mouse. Cell Rep. 10, 292–305. https://doi.org/10.1016/j.celrep.2014.12.014

Kirst, C., Skriabine, S., Vieites-Prado, A., Topilko, T., Bertin, P., Gerschenfeld, G., Verny, F., Topilko, P., Michalski, N., Tessier-Lavigne, M., Renier, N., 2020. Mapping the Fine-Scale Organization and Plasticity of the Brain Vasculature. Cell 1–16. https://doi.org/10.1016/j.cell.2020.01.028

Klein, A., Andersson, J., Ardekani, B.A., Ashburner, J., Avants, B., Chiang, M.C., Christensen, G.E., Collins, D.L., Gee, J., Hellier, P., Song, J.H., Jenkinson, M., Lepage, C., Rueckert, D., Thompson, P., Vercauteren, T., Woods, R.P., Mann, J.J., Parsey, R. V., 2009. Evaluation of 14 nonlinear deformation algorithms applied to human brain MRI registration. Neuroimage 46, 786–802. https://doi.org/10.1016/j.neuroimage.2008.12.037

Klein, S., Staring, M., Murphy, K., Viergever, M.A., Pluim, J.P.W., 2010. Elastix: A toolbox for intensity-based medical image registration. IEEE Trans. Med. Imaging 29, 196–205. https://doi.org/10.1109/TMI.2009.2035616

Kobat, D., Horton, N.G., Xu, C., 2011. In vivo two-photon microscopy to 1.6-mm depth in mouse cortex. J. Biomed. Opt. 16, 106014. https://doi.org/10.1117/1.3646209

Kuan, L., Li, Y., Lau, C., Feng, D., Bernard, A., Sunkin, S.M., Zeng, H., Dang, C., Hawrylycz, M., Ng, L., 2015. Neuroinformatics of the allen mouse brain connectivity atlas. Methods 73, 4–17. https://doi.org/10.1016/j.ymeth.2014.12.013

Kuwajima, T., Sitko, A., Bhansali, P., Jurgens, C., Guido, W., Mason, C., 2013. ClearT: a detergent- and solvent-free clearing method for neuronal and non-neuronal tissue. Development 140, 1364–1368. https://doi.org/10.1242/dev.091844

Lauterbur, P., 1973. Image Formation by Induced Local Interactions: Examples Employing Nuclear Magnetic Resonance. Nature 242, 190–191.

Lecun, Y., Bengio, Y., Hinton, G., 2015. Deep learning. Nature. https://doi.org/10.1038/nature14539

Lerner, T.N., Shilyansky, C., Davidson, T.J., Evans, K.E., Beier, K.T., Zalocusky, K.A., Crow, A.K., Malenka, R.C., Luo, L., Tomer, R., Deisseroth, K., 2015. Intact-Brain Analyses Reveal Distinct Information Carried by SNc Dopamine Subcircuits. Cell 162, 635–647. https://doi.org/10.1016/j.cell.2015.07.014

Liebmann, T., Renier, N., Bettayeb, K., Greengard, P., Tessier-Lavigne, M., Flajolet, M., 2016. Three-Dimensional Study of Alzheimer's Disease Hallmarks Using the iDISCO Clearing Method. Cell Rep. 16, 1138–1152. https://doi.org/10.1016/j.celrep.2016.06.060

Liu, L.D., Chen, S., Economo, M.N., Li, N., Svoboda, K., 2020. Accurate localization of linear probe electrodes across multiple brains. bioRxiv.https://doi.org/10.1101/2020.02.25.965210

Luzzati, F., Fasolo, A., Peretto, P., 2011. Combining confocal laser scanning microscopy with serial section reconstruction in the study of adult neurogenesis. Front. Neurosci. 5, 1–14. https://doi.org/10.3389/fnins.2011.00070

Majka, P., Bednarek, S., Chan, J., Jermakow, N., Liu, C., Saworska, G., Worthy, K., Silva, A., Wójcik, D., Rosa, M., 2020. Histology-based average template of the marmoset cortex with probabilistic localization of cytoarchitectural areas. bioRxiv. https://doi.org/10.1101/2020.04.10.036632

Mano, T., Murata, K., Kon, K., Shimizu, C., Ono, H., Yamada, R.G., Miyamichi, K., Susaki, E.A., Ueda, H.R., 2020. CUBIC-Cloud : An Integrative Computational Framework Towards Community-driven Whole-Mouse-Brain Mapping. bioRxiv https://doi.org/10.1101/2020.08.28.271031

Marstal, K., Berendsen, F., … M.S.-P. of the I., 2016, U., 2016. SimpleElastix: A user-friendly, multi-lingual library for medical image registration registration segmentation sensing unread, in: IEEE Conference on Computer Vision and Pattern Recognition Workshops (CVPRW),.

McQuin, C., Goodman, A., Chernyshev, V., Kamentsky, L., Cimini, B.A., Karhohs, K.W., Doan, M., Ding, L., Rafelski, S.M., Thirstrup, D., Wiegraebe, W., Singh, S., Becker, T., Caicedo, J.C., Carpenter, A.E., 2018. CellProfiler 3.0: Next-generation image processing for biology. PLoS Biol. 16, 1–17. https://doi.org/10.1371/journal.pbio.2005970

Menegas, W., Bergan, J. F., Ogawa, S. K., Isogai, Y., Venkataraju, K. U., Osten, P., Uchida, N., Watabe-Uchida, M. 2015. Dopamine neurons projecting to the posterior striatum form an anatomically distinct subclass. eLife 2015;4:e10032

Mehta, R., Majumdar, A., Sivaswamy, J., 2017. BrainSegNet: a convolutional neural network architecture for automated segmentation of human brain structures. J. Med. Imaging 4, 024003. https://doi.org/10.1117/1.jmi.4.2.024003

Minsky, M., 1961. Microscopy apparatus (US Patent 301467A). US 301467.

Modat, M., Ridgway, G.R., Taylor, Z.A., Lehmann, M., Barnes, J., Hawkes, D.J., Fox, N.C., Ourselin, S., 2010. Fast free-form deformation using graphics processing units. Comput. Methods Programs Biomed. 98, 278–284. https://doi.org/10.1016/j.cmpb.2009.09.002

Murakami, T.C., Mano, T., Saikawa, S., Horiguchi, S.A., Shigeta, D., Baba, K., Sekiya, H., Shimizu, Y., Tanaka, K.F., Kiyonari, H., Iino, M., Mochizuki, H., Tainaka, K., Ueda, H.R., 2018. A three-dimensional single-cell-resolution whole-brain atlas using CUBIC-X expansion microscopy and tissue clearing. Nat. Neurosci. https://doi.org/10.1038/s41593-018-0109-1

Myers, P.E., Arvapalli, G.C., Ramachandran, S.C., Pisner, D.A., Frank, P.F., Lemmer, A.D., Bridgeford, E.W., Nikolaidis, A., Vogelstein, J.T., 2019. Standardizing human brain parcellations. bioRxiv. https://doi.org/10.1101/845065

Ni, H., Feng, Z., Guan, Y., Jia, X., Chen, W., Jiang, T., Zhong, Q., Yuan, J., Ren, M., Li, X., Gong, H., Luo, Q., Li, A., 2020. DeepMapi: a Fully Automatic Registration Method for Mesoscopic Optical Brain Images Using Convolutional Neural Networks. Neuroinformatics. https://doi.org/10.1007/s12021-020-09483-7

Niedworok, C.J., Brown, A.P.Y., Jorge Cardoso, M., Osten, P., Ourselin, S., Modat, M., Margrie, T.W., 2016. AMAP is a validated pipeline for registration and segmentation of high-resolution mouse brain data. Nat. Commun. 7, 1–9. https://doi.org/10.1038/ncomms11879

Ogawa, S.K., Cohen, J.Y., Hwang, D., Uchida, N., Watabe-Uchida, M., 2014. Organization of monosynaptic inputs to the serotonin and dopamine neuromodulatory systems. Cell Rep. 8, 1105–1118. https://doi.org/10.1016/j.celrep.2014.06.042

Oh, S.W., Harris, J. a, Ng, L., Winslow, B., Cain, N., Mihalas, S., Wang, Q., Lau, C., Kuan, L., Henry, A.M., Mortrud, M.T., Ouellette, B., Nguyen, T.N., Sorensen, S. a, Slaughterbeck, C.R., Wakeman, W., Li, Y., Feng, D., Ho, A., Nicholas, E., Hirokawa, K.E., Bohn, P., Joines, K.M., Peng, H., Hawrylycz, M.J., Phillips, J.W., Hohmann, J.G., Wohnoutka, P., Gerfen, C.R., Koch, C., Bernard, A., Dang, C., Jones, A.R., Zeng, H., 2014. A mesoscale connectome of the mouse brain. Nature 508, 207–214. https://doi.org/10.1038/nature13186

Ortiz, C., Navarro, J.F., Jurek, A., Märtin, A., Lundeberg, J., Meletis, K., 2020. Molecular atlas of the adult mouse brain. Sci. Adv. 6, 1–14. https://doi.org/10.1126/sciadv.abb3446

Osten, P., Margrie, T.W., 2013. Mapping brain circuitry with a light microscope. Nat. Methods 10, 515–23. https://doi.org/10.1038/nmeth.2477

Papp, E.A., Leergaard, T.B., Calabrese, E., Johnson, G.A., Bjaalie, J.G., 2014. Waxholm Space atlas of the Sprague Dawley rat brain. Neuroimage 97, 374–386. https://doi.org/10.1016/j.neuroimage.2014.04.001

Perens, J., Gravesen, S.C., Skytte, J.L., Roostalu, U., Dahl, A.B., Dyrby, T.B., Wichern, F., Barkholt, P., Vrang, N., Jelsing, J., Hecksher-Sørensen, J., 2020. An Optimized Mouse Brain Atlas for Automated Mapping and Quantification of Neuronal Activity Using iDISCO+ and Light Sheet Fluorescence Microscopy. Neuroinformatics in press.

Pietzsch, T., Preibisch, S., Tomančák, P., Saalfeld, S., 2012. Img lib 2-generic image processing in Java. Bioinformatics 28, 3009–3011. https://doi.org/10.1093/bioinformatics/bts543

Pietzsch, T., Saalfeld, S., Preibisch, S., Tomancak, P., 2015. BigDataViewer: Visualization and processing for large image data sets. Nat. Methods 12, 481–483. https://doi.org/10.1038/nmeth.3392

Ragan, T., Kadiri, L.R., Venkataraju, K.U., Bahlmann, K., Sutin, J., Taranda, J., Arganda-Carreras, I., Kim, Y., Seung, H.S., Osten, P., 2012. Serial two-photon tomography for automated ex vivo mouse brain imaging. Nat. Methods 9, 255–258. https://doi.org/10.1038/nmeth.1854

Renier, N., Adams, E.L., Kirst, C., Wu, Z., Azevedo, R., Kohl, J., Autry, A.E., Kadiri, L., Umadevi Venkataraju, K., Zhou, Y., Wang, V.X., Tang, C.Y., Olsen, O., Dulac, C., Osten, P., Tessier-Lavigne, M., 2016. Mapping of Brain Activity by Automated Volume Analysis of Immediate Early Genes. Cell 165, 1789–1802. https://doi.org/10.1016/j.cell.2016.05.007

Renier, N., Wu, Z., Simon, D.J., Yang, J., Ariel, P., Tessier-Lavigne, M., 2014. iDISCO: A Simple, Rapid Method to Immunolabel Large Tissue Samples for Volume Imaging. Cell 159, 896–910. https://doi.org/10.1016/j.cell.2014.10.010

Ronneberger, O., Fischer, P., Brox, T., 2015. U-net: Convolutional networks for biomedical image segmentation, in: ArXiv. https://doi.org/10.1007/978-3-319-24574-4_28

Rust, M.J., Bates, M., Zhuang, X., 2006. Sub-diffraction-limit imaging by stochastic optical reconstruction microscopy (STORM). Nat. Methods 3, 793–796. https://doi.org/10.1038/nmeth929

Sage, D., Pham, T.A., Babcock, H., Lukes, T., Pengo, T., Chao, J., Velmurugan, R., Herbert, A., Agrawal, A., Colabrese, S., Wheeler, A., Archetti, A., Rieger, B., Ober, R., Hagen, G.M., Sibarita, J.B., Ries, J., Henriques, R., Unser, M., Holden, S., 2019. Super-resolution fight club: assessment of 2D and 3D single-molecule localization microscopy software. Nat. Methods 16, 387–395. https://doi.org/10.1038/s41592-019-0364-4

Schindelin, J., Arganda-Carreras, I., Frise, E., Kaynig, V., Longair, M., Pietzsch, T., Preibisch, S., Rueden, C., Saalfeld, S., Schmid, B., Tinevez, J.-Y., White, D.J., Hartenstein, V., Eliceiri, K., Tomancak, P., Cardona, A., 2012. Fiji: an open-source platform for biological-image analysis. Nat. Methods 9, 676–682. https://doi.org/10.1038/nmeth.2019

Seiriki, K., Kasai, A., Hashimoto, T., Schulze, W., Niu, M., Yamaguchi, S., Nakazawa, T., Inoue, K. ichi, Uezono, S., Takada, M., Naka, Y., Igarashi, H., Tanuma, M., Waschek, J.A., Ago, Y., Tanaka, K.F., Hayata-Takano, A., Nagayasu, K., Shintani, N., Hashimoto, R., Kunii, Y., Hino, M., Matsumoto, J., Yabe, H., Nagai, T., Fujita, K., Matsuda, T., Takuma, K., Baba, A., Hashimoto, H., 2017. High-Speed and Scalable Whole-Brain Imaging in Rodents and Primates. Neuron 94, 1085-1100.e6. https://doi.org/10.1016/j.neuron.2017.05.017

Siedentopf, H., Zsigmondy, R., 1903. Uber Sichtbarmachung und Größenbestimmung ultramikoskopischer Teilchen, mit besonderer Anwendung auf Goldrubingläser. Ann. Phys. 10, 1–39.

Skibbe, H., Watakabe, A., Nakae, K., Gutierrez, C.E., Tsukada, H., Hata, J., Kawase, T., Gong, R., Woodward, A., Doya, K., Okano, H., Yamamori, T., Ishii, S., 2019. MarmoNet: A pipeline for automated projection mapping of the common marmoset brain from whole-brain serial two-photon tomography. arXiv.

Sofroniew, N., Lambert, T., Evans, K., Nunez-Iglesias, J., Winston, P., Yamauchi, K., Bokota, G., Solak, A.C., ziyangczi, Buckley, G., Pop, D.D., Tung, T., Hector, Freeman, J., Bussonnier, M., Boone, P., Hilsenstein, V., Royer, L., Har-Gil, H., Lowe, A.R., Kittisopikul, M., Axelrod, S., alisterburt, Patil, A., McGovern, A., Rokem, A., Bryant, Gohlke, C., Kiggins, J., Huang, M., 2020. napari/napari: 0.4.1. https://doi.org/10.5281/ZENODO.4289401

Song, J.H., Choi, W., Song, Y.H., Kim, J.H., Jeong, D., Lee, S.H., Paik, S.B., 2020. Precise Mapping of Single Neurons by Calibrated 3D Reconstruction of Brain Slices Reveals Topographic Projection in Mouse Visual Cortex. Cell Rep. 31, 107682. https://doi.org/10.1016/j.celrep.2020.107682

Ståhl, P.L., Salmén, F., Vickovic, S., Lundmark, A., Navarro, J.F., Magnusson, J., Giacomello, S., Asp, M., Westholm, J.O., Huss, M., Mollbrink, A., Linnarsson, S., Codeluppi, S., Borg, Å., Pontén, F., Costea, P.I., Sahlén, P., Mulder, J., Bergmann, O., Lundeberg, J., Frisén, J., 2016. Visualization and analysis of gene expression in tissue sections by spatial transcriptomics. Science (80-. ). 353, 78–82. https://doi.org/10.1126/science.aaf2403

Stefaniuk, M., Gualda, E.J., Pawlowska, M., Legutko, D., Matryba, P., Koza, P., Konopka, W., Owczarek, D., Wawrzyniak, M., Loza-Alvarez, P., Kaczmarek, L., 2016. Light-sheet microscopy imaging of a whole cleared rat brain with Thy1-GFP transgene. Sci. Rep. 6, 1–9. https://doi.org/10.1038/srep28209

Susaki, E. a, Tainaka, K., Perrin, D., Kishino, F., Tawara, T., Watanabe, T.M., Yokoyama, C., Onoe, H., Eguchi, M., Yamaguchi, S., Abe, T., Kiyonari, H., Shimizu, Y., Miyawaki, A., Yokota, H., Ueda, H.R., 2014. Whole-brain imaging with single-cell resolution using chemical cocktails and computational analysis. Cell 157, 726–39. https://doi.org/10.1016/j.cell.2014.03.042

Susaki, E.A., Shimizu, C., Kuno, A., Tainaka, K., Li, X., Nishi, K., Morishima, K., Ono, H., Ode, K.L., Saeki, Y., Miyamichi, K., Isa, K., Yokoyama, C., Kitaura, H., Ikemura, M., Ushiku, T., Shimizu, Y., Saito, T., Saido, T.C., Fukayama, M., Onoe, H., Touhara, K., Isa, T., Kakita, A., Shibayama, M., Ueda, H.R., 2020. Versatile whole-organ/body staining and imaging based on electrolyte-gel properties of biological tissues. Nat. Commun. 11. https://doi.org/10.1038/s41467-020-15906-5

Theer, P., Hasan, M.T., Denk, W., 2003. Two-photon imaging to a depth of 1000 microm in living brains by use of a Ti:Al2O3 regenerative amplifier. Opt. Lett. 28, 1022–1024. https://doi.org/10.1364/OL.28.001022

Todorov, M.I., Paetzold, J.C., Schoppe, O., Tetteh, G., Shit, S., Efremov, V., Todorov-Völgyi, K., Düring, M., Dichgans, M., Piraud, M., Menze, B., Ertürk, A., 2020. Machine learning analysis of whole mouse brain vasculature. Nat. Methods 17, 442–449. https://doi.org/10.1038/s41592-020-0792-1

Tomer, R., Ye, L., Hsueh, B., Deisseroth, K., 2014. Advanced CLARITY for rapid and high-resolution imaging of intact tissues. Nat. Protoc. 9, 1682–97. https://doi.org/10.1038/nprot.2014.123

Tward, D., Li, X., Huo, B., Lee, B., Miller, M., Mitra, P., 2020. Solving the where problem in neuroanatomy: a generative framework with learned mappings to register multimodal, incomplete data into a reference brain. nbio. https://doi.org/10.1101/2020.03.22.002618

Tyson, A. L., Rousseau, C. V., Margrie, T.W., 2020. brainreg: automated 3D brain registration with support for multiple species and atlases. https://doi.org/10.5281/ZENODO.3991718

Tyson, A. L, Rousseau, C. V, Niedworok, C.J., Keshavarzi, S., Tsitoura, C., Margrie, T.W., 2020. A deep learning algorithm for 3D cell detection in whole mouse brain image datasets. Biorxiv https://doi.org/10.1101/2020.10.21.348771

Ulman, V., Maška, M., Magnusson, K.E.G., Ronneberger, O., Haubold, C., Harder, N., Matula, Pavel, Matula, Petr, Svoboda, D., Radojevic, M., Smal, I., Rohr, K., Jaldén, J., Blau, H.M., Dzyubachyk, O., Lelieveldt, B., Xiao, P., Li, Y., Cho, S.Y., Dufour, A.C., Olivo-Marin, J.C., Reyes-Aldasoro, C.C., Solis-Lemus, J.A., Bensch, R., Brox, T., Stegmaier, J., Mikut, R., Wolf, S., Hamprecht, F.A., Esteves, T., Quelhas, P., Demirel, Ö., Malmström, L., Jug, F., Tomancak, P., Meijering, E., Muñoz-Barrutia, A., Kozubek, M., Ortiz-De-Solorzano, C., 2017. An objective comparison of cell-tracking algorithms. Nat. Methods 14, 1141–1152. https://doi.org/10.1038/nmeth.4473

Valdés-Hernández, P.A., Sumiyoshi, A., Nonaka, H., Haga, R., Aubert-Vásquez, E., Ogawa, T., Iturria-Medina, Y., Riera, J.J., Kawashima, R., 2011. An in vivo MRI Template Set for Morphometry, Tissue Segmentation, and fMRI Localization in Rats. Front. Neuroinform. 5, 26. https://doi.org/10.3389/fninf.2011.00026

Van Der Walt, S., Schönberger, J.L., Nunez-Iglesias, J., Boulogne, F., Warner, J.D., Yager, N., Gouillart, E., Yu, T., 2014. Scikit-image: Image processing in python. PeerJ 2014, 1–18. https://doi.org/10.7717/peerj.453

Vélez-Fort, M., Rousseau, C. V, Niedworok, C.J., Wickersham, I.R., Rancz, E.A., Brown, A.P.Y., Strom, M., Margrie, T.W., 2014. The stimulus selectivity and connectivity of layer six principal cells reveals cortical microcircuits underlying visual processing. Neuron 83, 1431–1443. https://doi.org/10.1016/j.neuron.2014.08.001

Voie, A., Burns, D., Spelman, F., 1993. Orthogonal-plane fluorescence optical sectioning: three-dimensional imaging of macroscopic biological specimens. J. Microsc. 170, 229–236.

Voigt, F.F., Kirschenbaum, D., Platonova, E., Pagès, S., Campbell, R.A.A., Kastli, R., Schaettin, M., Egolf, L., van der Bourg, A., Bethge, P., Haenraets, K., Frézel, N., Topilko, T., Perin, P., Hillier, D., Hildebrand, S., Schueth, A., Roebroeck, A., Roska, B., Stoeckli, E.T., Pizzala, R., Renier, N., Zeilhofer, H.U., Karayannis, T., Ziegler, U., Batti, L., Holtmaat, A., Lüscher, C., Aguzzi, A., Helmchen, F., 2019. The mesoSPIM initiative: open-source light-sheet microscopes for imaging cleared tissue. Nat. Methods 16, 1105–1106. https://doi.org/10.1038/s41592-019-0554-0

Wan, P., Zhu, J., Xu, J., Li, Y., Yu, T., Zhu, D., 2018. Evaluation of seven optical clearing methods in mouse brain. Neurophotonics 5, 1. https://doi.org/10.1117/1.nph.5.3.035007

Wang, Q., Ding, S.L., Li, Y., Royall, J., Feng, D., Lesnar, P., Graddis, N., Naeemi, M., Facer, B., Ho, A., Dolbeare, T., Blanchard, B., Dee, N., Wakeman, W., Hirokawa, K.E., Szafer, A., Sunkin, S.M., Oh, S.W., Bernard, A., Phillips, J.W., Hawrylycz, M., Koch, C., Zeng, H., Harris, J.A., Ng, L., 2020. The Allen Mouse Brain Common Coordinate Framework: A 3D Reference Atlas. Cell 181, 936-953.e20. https://doi.org/10.1016/j.cell.2020.04.007

Watabe-Uchida, M., Zhu, L., Ogawa, S.K., Vamanrao, A., Uchida, N., 2012. Whole-Brain Mapping of Direct Inputs to Midbrain Dopamine Neurons. Neuron 74, 858–873. https://doi.org/10.1016/j.neuron.2012.03.017

Winnubst, J., Bas, E., Ferreira, T.A., Wu, Z., Economo, M.N., Edson, P., Arthur, B.J., Bruns, C., Rokicki, K., Schauder, D., Olbris, D.J., Murphy, S.D., Ackerman, D.G., Arshadi, C., Baldwin, P., Blake, R., Elsayed, A., Hasan, M., Ramirez, D., Dos Santos, B., Weldon, M., Zafar, A., Dudman, J.T., Gerfen, C.R., Hantman, A.W., Korff, W., Sternson, S.M., Spruston, N., Svoboda, K., Chandrashekar, J., 2019. Reconstruction of 1,000 Projection Neurons Reveals New Cell Types and Organization of Long-Range Connectivity in the Mouse Brain. Cell 179, 268-281.e13. https://doi.org/10.1016/j.cell.2019.07.042

Woodward, A., Hashikawa, T., Maeda, M., Kaneko, T., Hikishima, K., Iriki, A., Okano, H., Yamaguchi, Y., 2018. Data descriptor: The Brain/MINDS 3D digital marmoset brain atlas. Sci. Data 5, 1–12. https://doi.org/10.1038/sdata.2018.9

Young, D.M., Duhn, C., Gilson, M., Nojima, M., Yuruk, D., Kumar, A., Yu, W., Sanders, S.J., 2020. Whole-Brain Image Analysis and Anatomical Atlas 3D Generation Using MagellanMapper. Curr. Protoc. Neurosci. 94, e104. https://doi.org/10.1002/cpns.104